# Computer Art in the (Former) Soviet Bloc

Eric Engle

**Introduction**

The development of computer media occurred in Eastern Europe as the result of a logical progression from still photographs, to film, to computer generated still images to computer generated moving images. Thus video art is the proper precursor to interactive art.[1] These tendencies must be understood in their material, economic, historic, and technological dimensions.[2]

**I. Experimental visual art in Yugoslavia**

Prior to exposing the technical achievements and political failure of Yugoslavian computer art we must understand some basic terms. Otherwise we will not be able to place the technical achievements of Yugoslavian computer art into the larger context of the political failure of the Yugoslavian regime.

*A. The Maoist Critique of Yugoslavia*

**1. Revolutionary Definitions**

In order to understand the Maoist critique of Tito and Yugoslavia it is necessary to understand some basic terms. When scientists use a term they mean something fairly precise. Here the terms we need to define are "revisionism" and "bourgeoisie". We could look at other terms as well – "opportunism", "fascism", "sub-reformism", for Maoists assign very specific meanings to those terms as well. However the heart of the problem of Yugoslavia can be addressed with just these two terms.

A marxist glossary provides the following definitions:

*Revisionism:* "Revisionism refers to political views that claim to be Marxist yet revise Marx's work fundamentally by failing to apply the scientific method of dialectical materialism. Revisionists commonly downplay class struggle, overplay the struggle to increase production and technical progress compared with political matters, don't believe imperialism is dangerous, advocate reformist means of change and don't uphold the dictatorship of the proletariat. MIM also calls revisionists phoney communists or state capitalists if they are in power."[3]

*Labor aristocracy.* "'Labor aristocracy' refers to the working class that benefits from the imperialist world's super-exploitation of the Third World. For example, white workers in the United States benefit from the super-exploitation of the Third World so greatly, that as a class, they are no longer exploited at all and in fact benefit from imperialism."[4]

The final term we need to understand to properly assess art in Eastern Europe during the cold war is "*capitalist restoration*". Mao explained that socialism can be defeated by capitalism through the rise of a bourgeoisie within the communist party because of corruption. Ultimately this leads to a capitalist restoration.[5] Capitalist restoration has happened in every nominally Marxist state. Revisionism eventually leads to capitalist restoration.[6] Essentially Yugoslavia suffered capitalist restoration due to revisionism resulting in fascism and genocidal ethnic cleansing. Yugoslavias history proves Mao's theory was correct.

**2. Criticisms of Tito**

Essentially, Mao's critique of Tito and Yugoslavia – echoed by Stalin's third in Command V.M. Molotov later in the 1970s,[7] is that the Yugoslavian regime was revisionist.[8] That it was "cozying up to capitalism". This had an unintended secondary effect: by cozying up to capitalism and implementing local autonomy the nationalities in Yugoslavia were never integrated. These political tendencies were naturally reflected in video art,[9] and by extension to computer art as well since computer art developed out of video art.

History has proven Mao right. Capitalist restoration hit Yugoslavia and unleashed fascism – just as it did in East Germany and Russia. But in Yugoslavia he fascists backlash was even worse since there was no national unity – at least not within a federal Yugoslavia. Even the member states of the Yugoslavian federation had ethnic minorities. The result of the poisonous mixture of revisionism and ignorance of the nationality issue? Deadly ethnic cleansing.[10]

*B. New Tendencies*

We have already noted the technical competence of Eastern Europe in computer art generally. What is true of computer art in Eastern Europe generally is specifically true of Yugoslavia which, with East Germany, is one of the leading examples of computer art in Eastern Europe during the COMECON era. Yugoslavian art developed in a historical progression according to technologies availability. However, Yugoslavia, due to revisionism, had greater access to Western technology and personages.[11] Thus it was technically one of the leading countries in the Eastern bloc but politically one of the least advanced. Yugoslavian revisionism was most succesful at attracting western artists including computer artists to Yugoslavia. Thus, "In 1969 Zagreb was first after London to take part in an international congress on computer art "Dialogue with the Machine" (Dijalog sa Strojem). In 1971 "Television Today" (Televizija danas) edited by Vera Horvat-Pintaric, attention was already being paid to experimental forms of the mass media."[12] We look at Yugoslavia as an example of a combination of political failure and technical genius. However this technological achievement must be placed in its political context. The Yugoslavian Expirement in Market Socialism proved itself to be a failed strategy. Decentralization of the means of production rather than increasing local autonomy and freedom resulted in a rapid capitalist restoration. Worse, it also failed to address the question of different nations in the Yugoslav Federation. Consequently, the erroneous line of Yugoslavia set the stage for genocidal ethnic cleansing.

The crime of the failure of the Yugoslavian state to prevent capitalist restoration, fascism, and the resulting genocidal "ethnic cleansing" is made all the more poignant due to the fact that the Yugoslavian art scene was very creative. Because the Yugoslavian government was constantly trying to ingratiate itself with the West it had access to resources both human and material that the other nominally communist states of Eastern Europe did not have. The Yugoslavian computer-art scene can be seen through the lense of the "nove tendencije" – new tendencies movement.

These contradictory aspects of computer art in Yugoslavia can be best addressed by considering the "New Tendencies" art movement of the late 1960s and early 1970s.

The leading edge of the revisionist wave was known as "New Tendencies":

"At the beginning of the sixties the country opened its doors which had a vital influence on artistic creativity, music and short film. The time was ripe for a new approach to both art and criticism. New views were developed which stimulated thinking about the function, significance, place and role of art in modern society. As a part of the artist's movement "Nove tendencije", (New Trends) movement numerous events, exhibitions and festivals took place and Zagreb developed into one of the centres of modern art in Europe."[13]

Revisionism made "new tendencies" attractive to the capitalist countries. In their defence, these artists were working with new technologies. Certainly as individuals many of them were well intentioned. Some even have shown themselves committed to progressive causes, even at times despite their own class privelege. However as a group, nove tendencije seems to have only shaken the world up and not (yet) taken it by storm.

Whether the conservative tendencies in art after 1971[14] were due to anti-revisionism or, more likely, party bureaucrats defending the bourgeoisie which seems to have already emerged within the Yugoslavian communist party is an interesting question. However even if there were anti-revisionist elements in Yugoslavia they were not strong enough to prevent capitalist restoration or even to mitigate its worst consequences. However, from a technical perspective and also as a leading example of computer art in a nominally communist country the new tendencies movement deserves examination – if only to demonstrate that its art was basically devoid of revolutionary potential. However, while the class character of this art was disappointing the technical vision of the artists is at times nearly prophetic. This also justifies an extended examination of New Tendencies.

As a system of thought New Tendencies attempted to integrate several different artistic movements of the 1960s and 1970s.[15] The movement begain in 1961 and the work it produced was considered "tautological" and "monochromatic", for example:

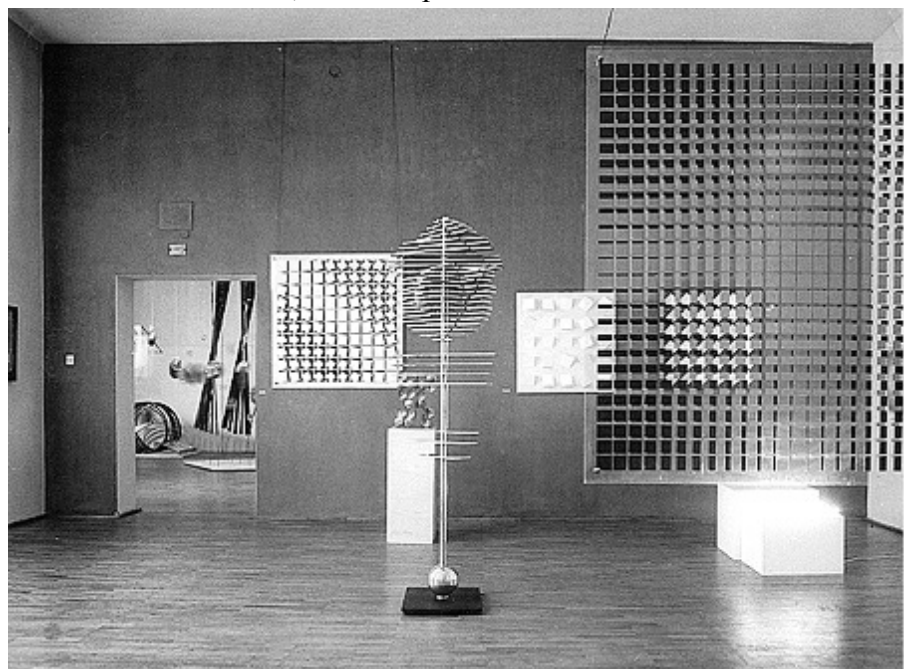

(Source: Museum of Contemporary Art, Zagreb, *New Tendencies*, http://www.mdc.hr/msu/media/slike/zbirke/223-inozumj/223-02-v.gif)

As can be seen, the results, artistically, are rather empty. One does not see a proletarian art which seeks, except perhaps in the glorification of the machine, to struggle for progress. Instead one sees algorithmic expression of algebraic formulae. This as a celebration of science certainly has some merit. But revolutionary imagery seems absent, which is unsurprising since revisionism meant that the idea of revolution was also absent from the Yugoslavian political landscape. The only hope is for radical revolutionary subtexts to be extracted by the watchful. But at least this viewer does not see them.

Thus, rather than resulting in radical critiques instead merely paved the way for capitalist restoration in Yugoslavia. Though Yugoslavia was willing to experiment with many new art forms and, thanks to revisionism, had access to them that did not at all prevent fascism and genocide. Thus in the "red" and "expert" debate it seems clear that the political line should dominate over expertise. To place technology, and not politics, in command led to needless deaths in Yugoslavia due to the restoration of capitalism.

On the other hand, though apparently politically void, New Tendencies was focused on technical progress: but systematic and working towards optical research.[16] Kinetic art and Gestalt theory were part of new tendencies by at least 1963. In 1965 the third third exhibition of New Tendencies considered cybernetic art and resulted in a symposium on computers and art. This theme was treated again in the fourth New Tendencies exhibition in 1968/69 which looked at a theory of information and exact aesthetics was marked by a further penetration of the idea on theory of information and exact aesthetics. A magazine was also launched, "Bit international" (nos. 1-9/1968-1972).[17] It would only be possible to evaluate whether New Tendencies was definitively a political or aesthetic void by obtaining and analyzing their writings, which are unfortunately in Serbo-Croation. The fifth and final New Tendencies exhibition also addressed visual computational art as illustrated below:

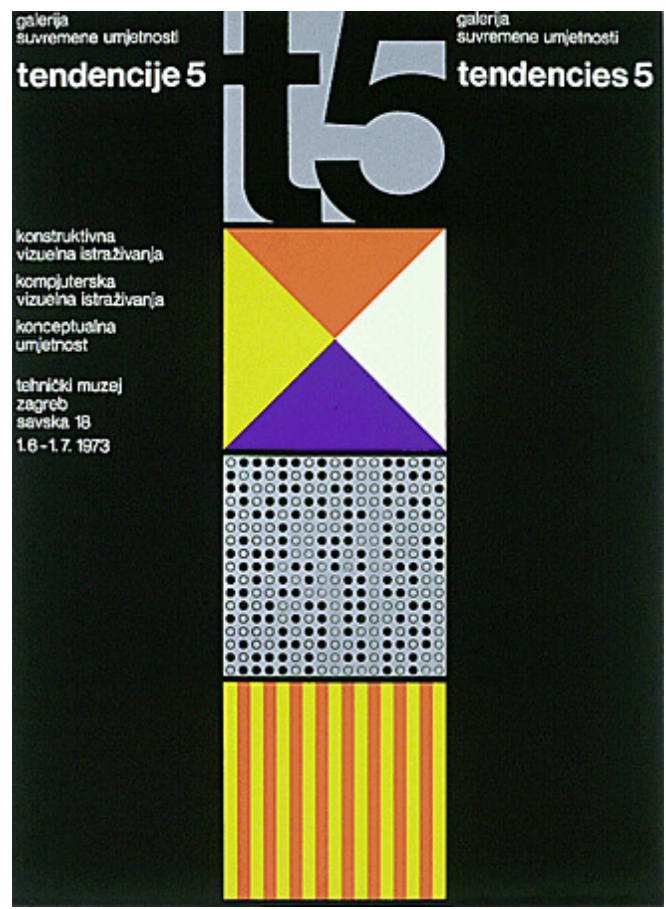

(Source: Museum of Contemporary Art, Zagreb, *New Tendencies*,

http://www.mdc.hr/msu/media/slike/zbirke/226-plak-dz/226-02-v.gif)

However, some snippets of thei aesthetic of New Tendencies are available. For example, Darko Fritz stated in *Amnesia International* that "... Technology progresses. Art changes. It never progresses. …" statement of the collective Anonima in May 1968, catalogue 'tendencije 4' (1968 - 69), Zagreb, 1970. That is a very reactionary statement on the one hand because it says that there is no social progress, that different productive modes are aesthetically equivalent. But we know for a fact that with historical progress social violence diminishes. On the other hand we could interpret this statement to mean that the same art which is beautiful in pre-agrarian primitive communist societies is just as beautiful now as then. Thus there would be an objective artistic aesthetic. With greater context we can try to isolate the potentially reactionary and potentially liberating tendencies in this seemingly innocuous phrase. Fritz goes on to say that

"It seems interesting to me that in the years 1968-69, amidst the Cold war, it was possible to bring together, under the title "Computers and Visual Research", the authors and theoreticians from both blocks (USA, USSR, Argentina, West and East Europe). Back then files probably did not occupy more than 1Mb, yet graphics, films, objects, sculptures modelled in 3D, music, choreography were present… Bonačić placed a large object *on permanent display* on the frontispiece of a department store in Zagreb. It seems to me that nowadays the possibility of choice and movement is incomparably greater … yet it is interesting to see the results … in the period 1968-72 nine thematic issues of the media art publication 'bit international' were published."[18]

While Fritz might believe that working to cooperate with capitalism is the best way to world peace the history of the last 10 years has proven him wrong. In fact, the collapse of the Soviet bloc has led to a worsening of human rights in South Africa, Russia and probably throughout the world as a whole given that the U.S. is openly torturing captives in Iraq and at Guantanamo bay. Again, we can see revisionism lurking here if we only know how to look for it.

New Tendencies was not only revisionism overt or covert. Regarding technology, the Anonima art collective stated that:

> ... pure technology is always more interesting and more beautiful than the art amalgamated with technology. … [*][19]

Again, this seems to be isolating technology from the culture that creates it with the result that we could ignore that the wealth used to generate technology in capitalism is derived from the exploitation of the third world. However New Tendenices also recognized that technology could be a force of alienation and.[20] They attempted to use technology to overcome technologies potential for alienation.[21] Thus some of the revisionist potential in New Tendencies while unrecognized was unconsciously resisted.

New Tendencies was also visionary. Though recognizing the obvious limitations of computing power in the late 1960s [22] in the Zagreb Manifesto they predicted the rise of the internet for example. [23] But lurking beneath prophetic dreams were revisionist tendencies clearly seen in the following passage:

> "the first award winners in the now annual computer art contest organised by the 'Computers and Automation' were the member of the U.S ballistic team. There is no doubt that in the computer art the real avant-garde was army."[24]

So at least some members of New Tendencies were, perhaps unwittingly, openly admitting the reactionary character of New Tendencies!

We can see the absence of a proletarian message when we look at the art created by New Tendencies. It seems focussed on technology and aesthetically neutral. However in a world where starvation, war and homelessness are a reality for the majority of the planet neutrality is not really neutral: it serves the interests of those who profit from poverty.

The various artefacts created by New Tendencies are available on-line at the Zagreb museum.[25] We present the ones which were computer generated or which use computer displays here.

## 1. NAMA DEPARTMENT STORE

One of the larger sculpture installation of New Tendencies, celebrating consumption, was a computer controlled electronic display at the Nama department store made by Vladimir Bonačić. This installation, "exclude[d] the chance by means of a pseudo-random polynomial to the 18th power exhibited on a 36 meter long series of 18 objects"[26] (top of photograph).

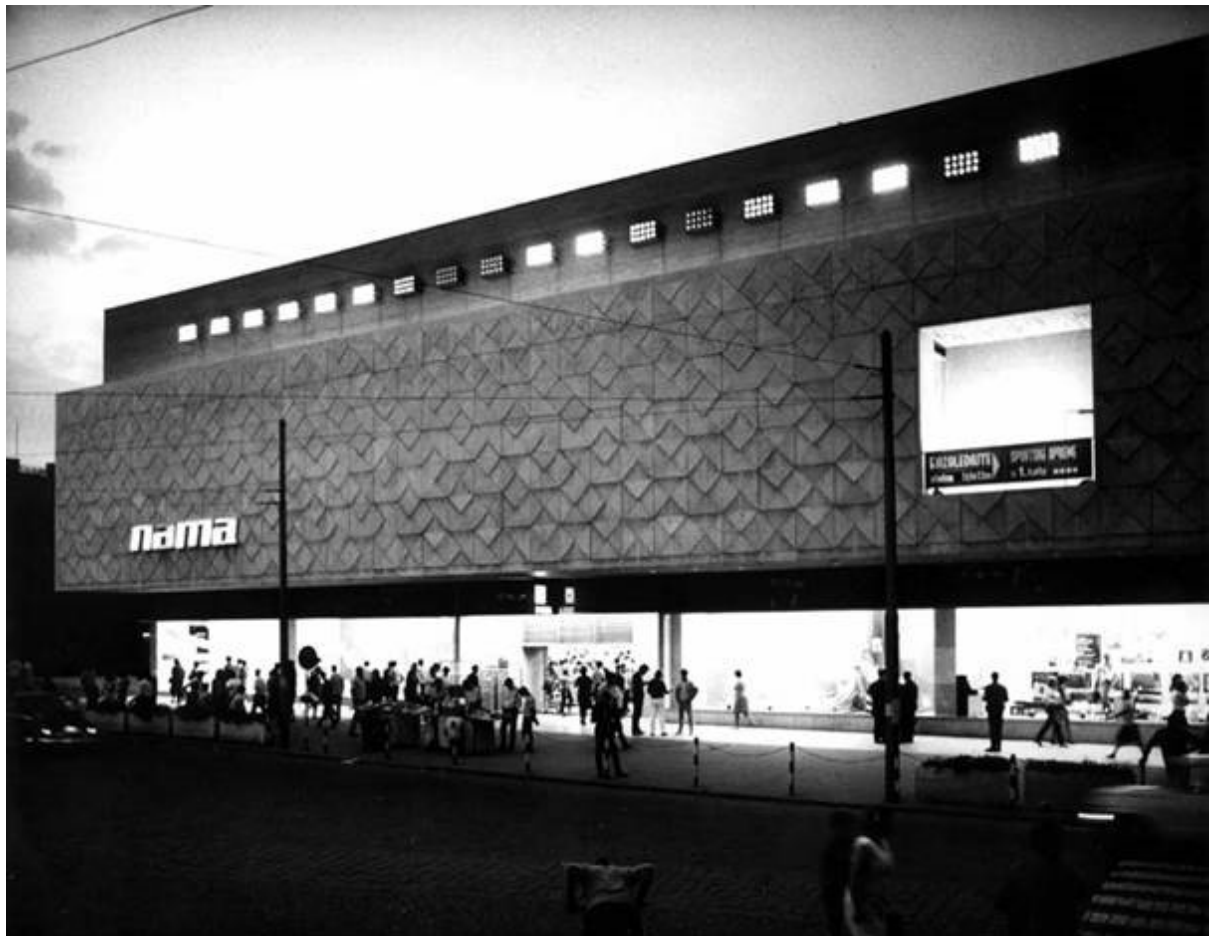

(Source: http://mama.mi2.hr/alive/eng/popis.htmhttp://mama.mi2.hr/alive/pix/mbonacicnama1.jpg)

## 2. Other Examples of New Tendencies

Other art created by New Tendencies was less spectacular but also focused on artistic expression of mathematical formulae. Several examples follow:

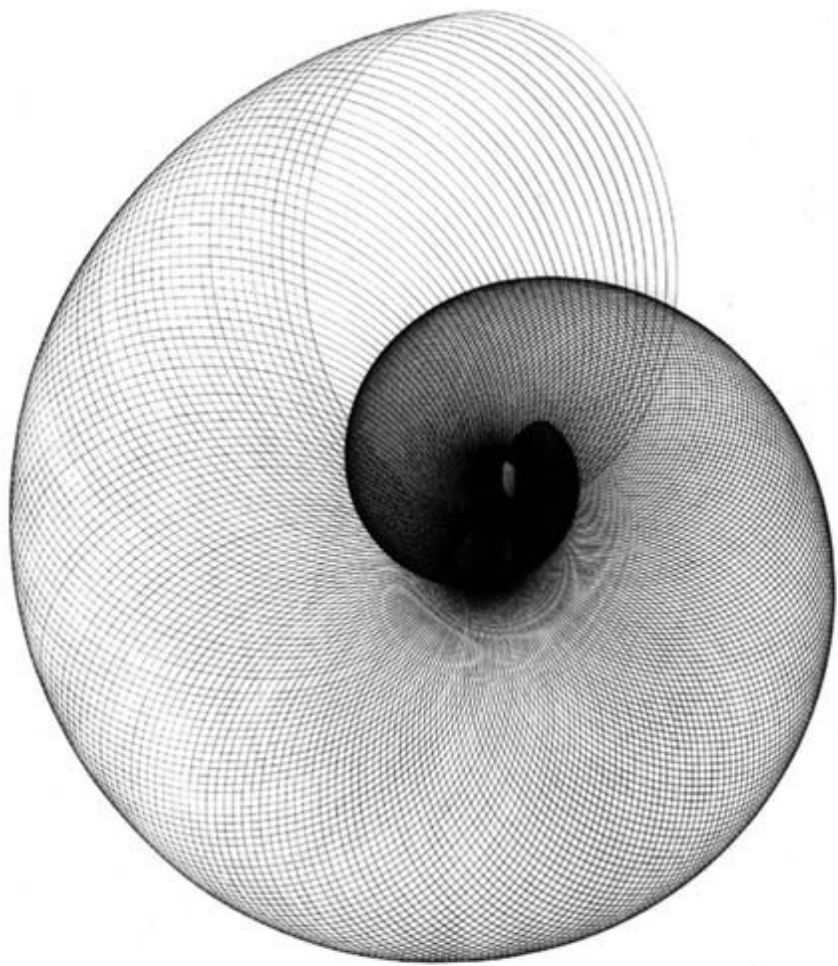

(Source: http://mama.mi2.hr/alive/pix/m05.france.jpg)

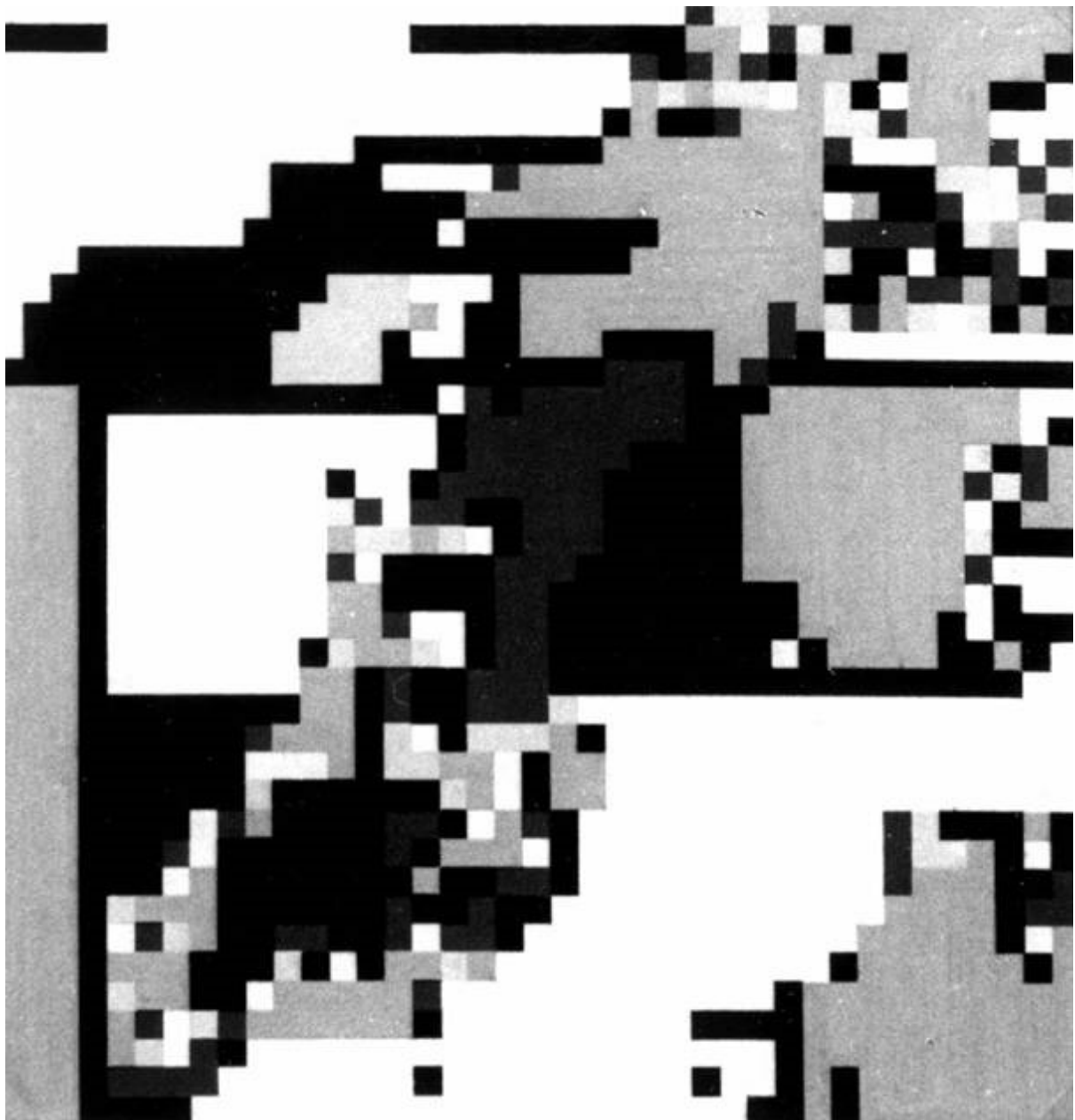

(Source: http://mama.mi2.hr/alive/pix/m06kawano.jpg)

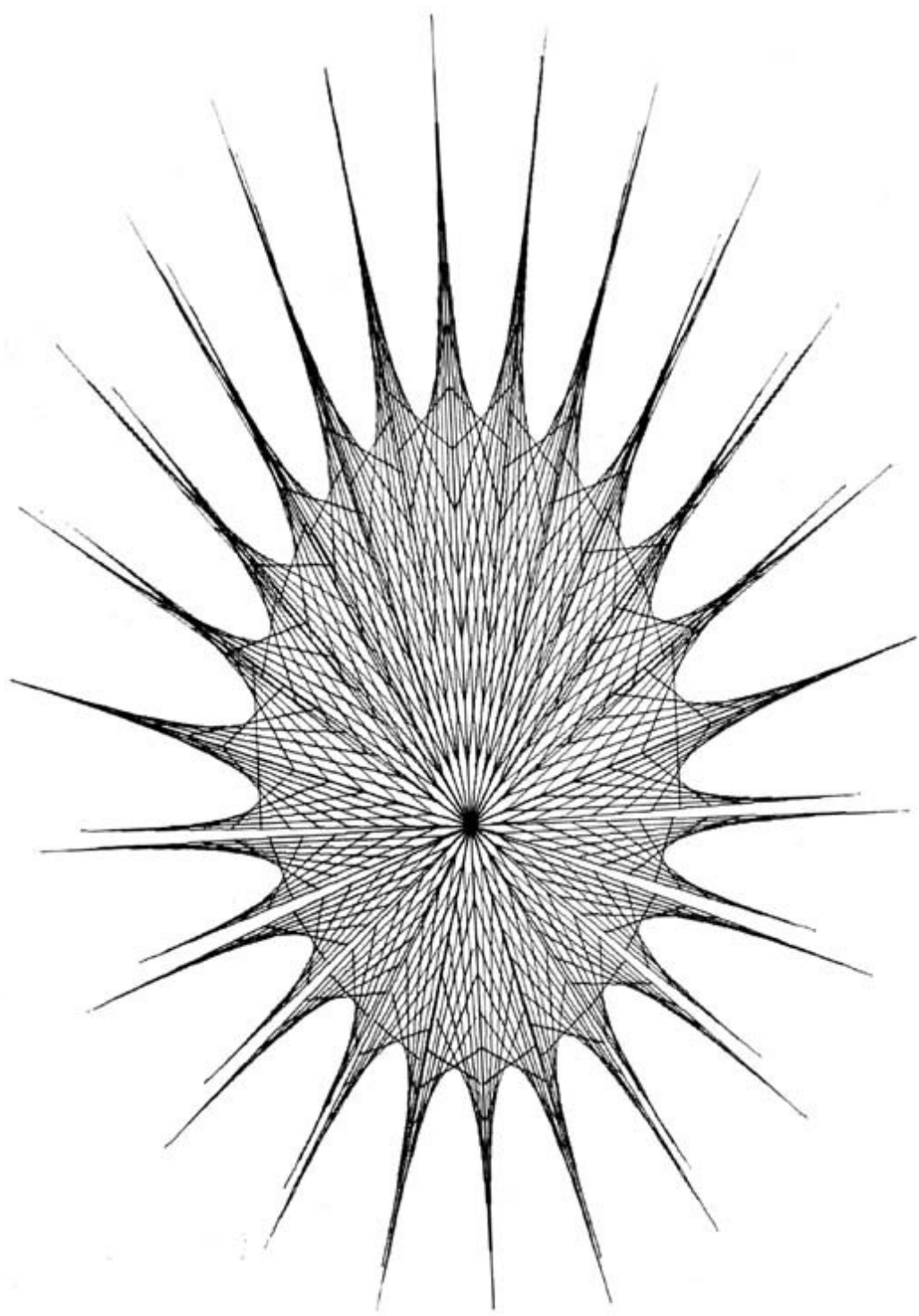

(Source: http://mama.mi2.hr/alive/pix/m12milojevic.jpg)

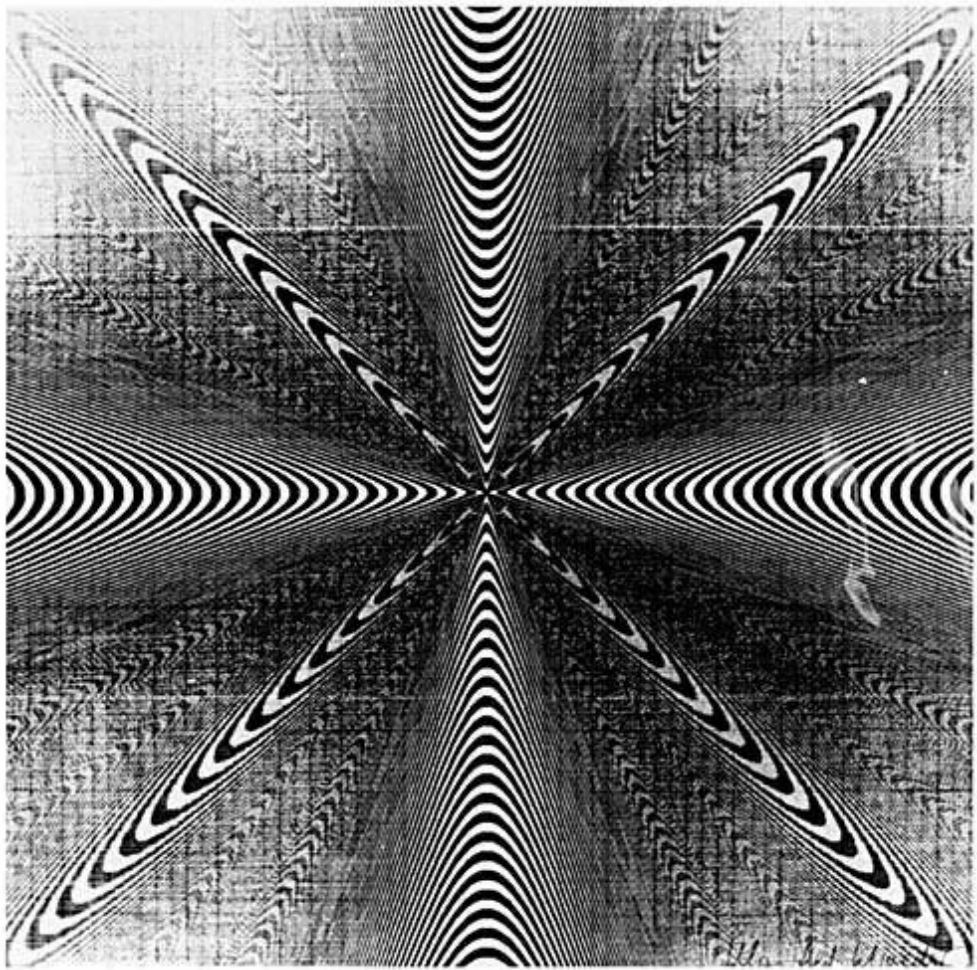

(Source: http://mama.mi2.hr/alive/pix/m27schroeder.jpg)

Nake

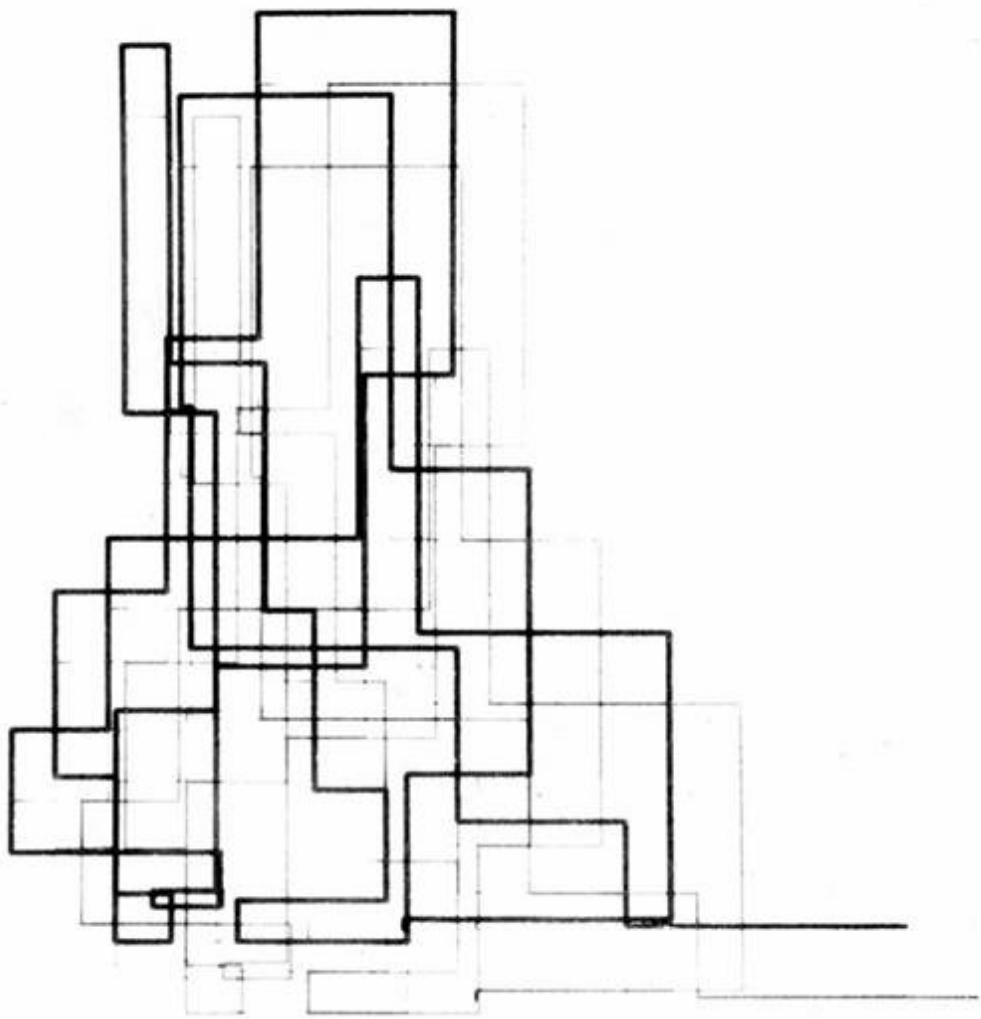

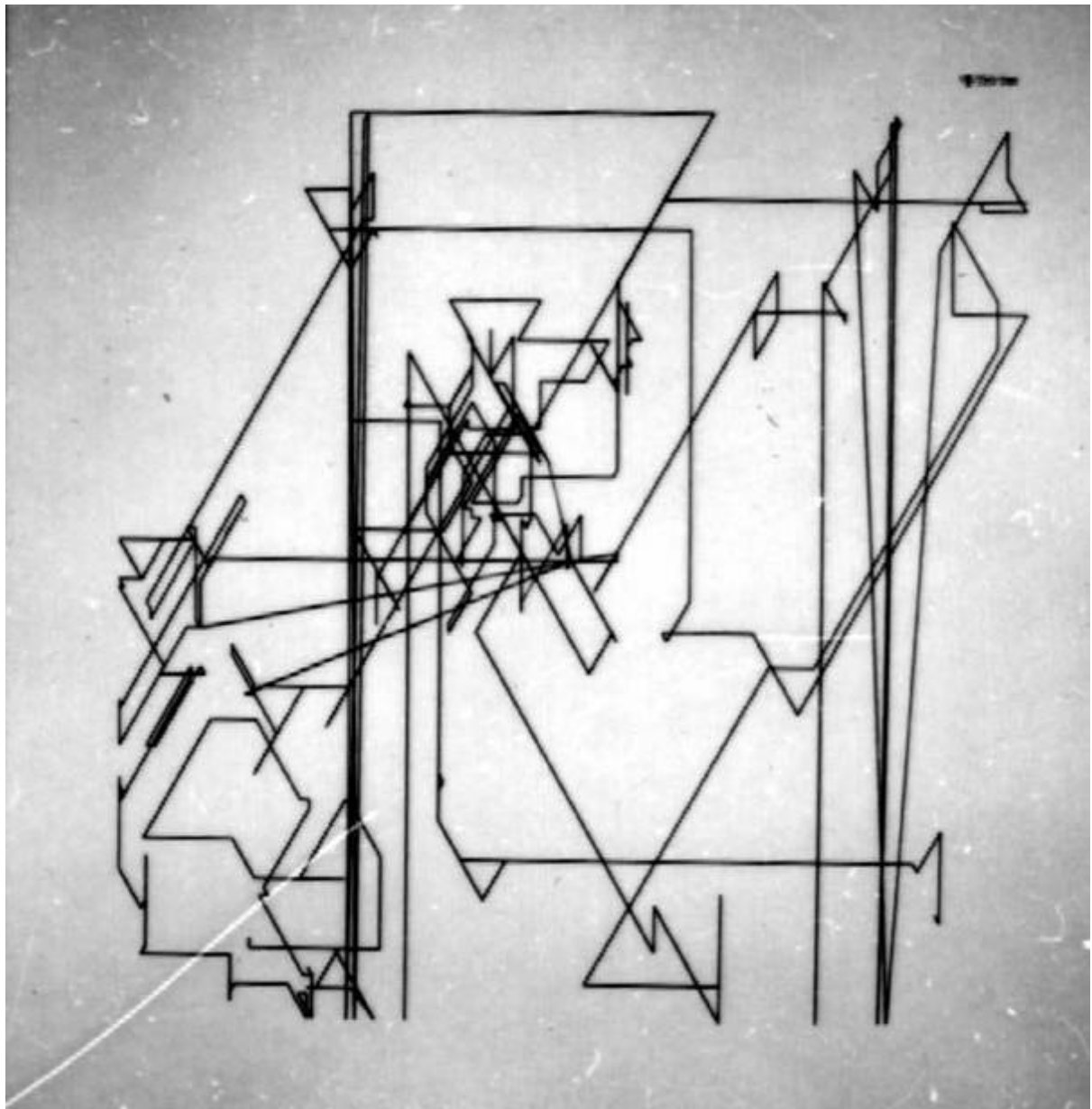

(Source: http://mama.mi2.hr/alive/pix/m15nake.jpg; http://mama.mi2.hr/alive/pix/m14nake.jpg)

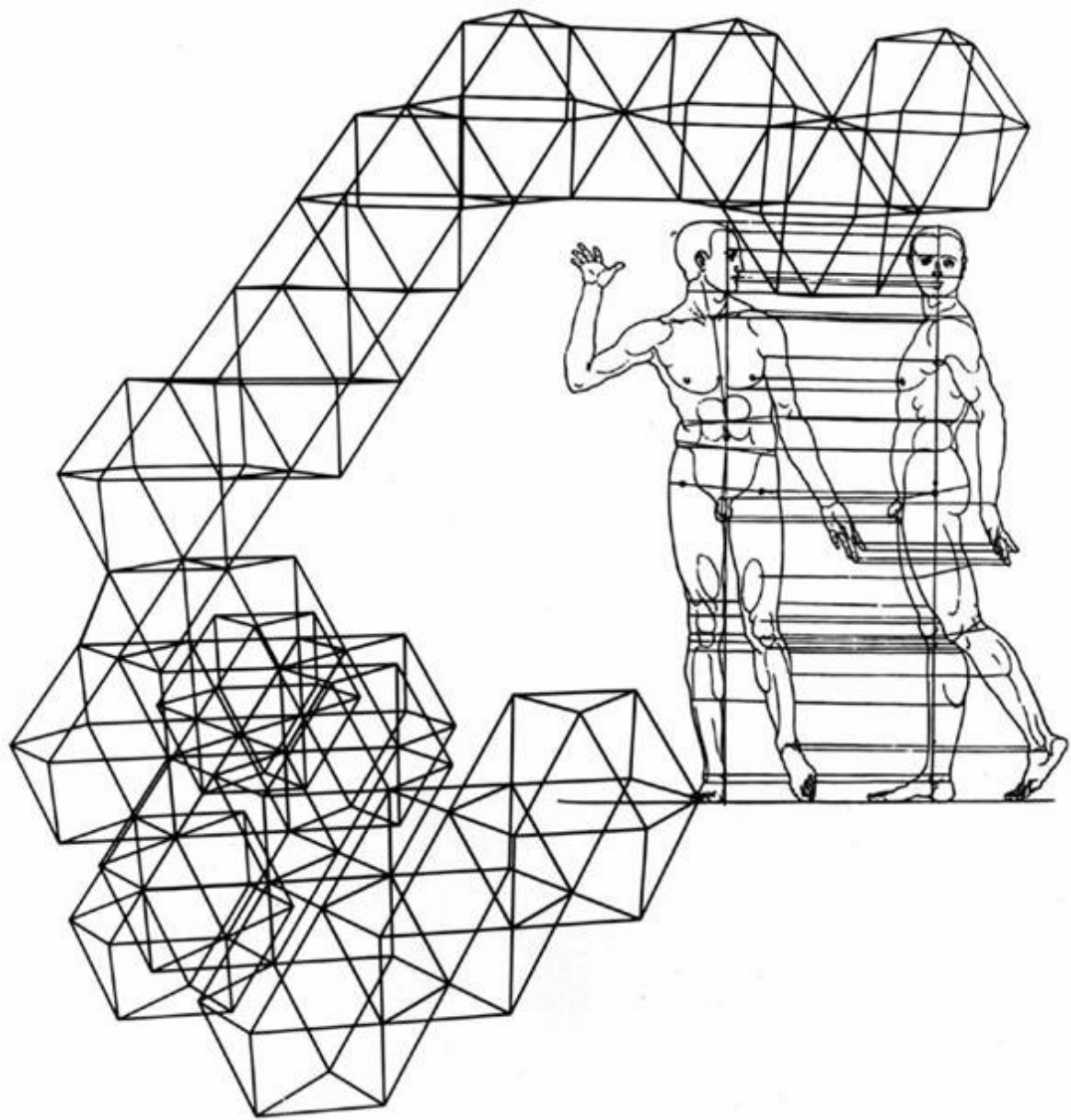

(Source: http://mama.mi2.hr/alive/pix/m26rasenees.jpg)

### C. The Successors to New Tendencies:

Though technically excellent, even visionary, New Tendencies seemed to lack political impetus, perhaps due to the erroneous idea that aesthetics are somehow neutral and that a "pure" art which ignores social conflict and that art is a social production could be possible. However there is at least a bit of hope: out of the successors of New Tendencies, one Vok Cusik has emerged. His art style is both refreshing and perhaps even revolutionary. He digitizes films into ASCII art. ASCII art films could require less bandwidth than "real ones" – allowing their broader diffusion to the poor. Also the topics he treats are tantalizing. He has digitized at least one revolutionary film, Eisenstein's "Potemkin".[27] Perhaps the generation suceeding "New Tendencies" will understand the political errors that undermined their artistic work.

### II. Computer art in East Germany

The story of East German computer art is rather happier than that of Yugoslavia. Though neo-fascists have arisen in the former East Germany they have not yet presented a coherent political force. In all events there was no genocide in East Germany – unlike Yugoslavia. During the nominally communist era East Germany was the most productive East European economy. It may surprise the reader, but East Germans had personal computers. They even used their computers to make art! Several interesting technological innovations arose out of the vibrant East German computer scene whose computers were manufactured by Robotron.

Politically, having suffered terribly under fascism, East Germany was also less subject to revisionism. That may seem paradoxical but the history seems to bear the statement out. The East German economy was the most productive of all Eastern Europe. East Germans enjoyed a lifestyle comparable with the poorer states of Western Europe, complete with automobiles, vacations  - and computers. This prosperity however did not deteriorate into decadence or revisionism. This – just as the Soviet space program and victory in the second world war -  shows that a worker's party guided by principles of Marxism rather than political opportunism can appreciably improve the well being of its citizens.

#### *A. Programme Funken!*

One interesting innovation of the East German computer industry was using radios to disseminate computer programs. Rather than disseminating the computer code, the file itself was transmitted for recording by the audience. At that time most computers used cassettes for storage. This innovation never occurred in the west because it would be unprofitable to merely give away computer programs.[28] Similarly, programs were also stored on vinyl records (unlike the west).[29] In technologies for data storage East Germany was a world leader. In other aspects of computer technology however it lagged roughly four to five years behind the West.[30]

#### *B. Computer Art in DDR*

Similarly, when we look at the software, we see that German Robotron computers included word processors, spreadsheets – and graphics programs: just like their western counterparts.[31]

Looking at the covers of popular East German programming books also reveals that graphic representation of mathematical formulas was just as fascinating to East Germans as to West Germans. These are illustrated below.

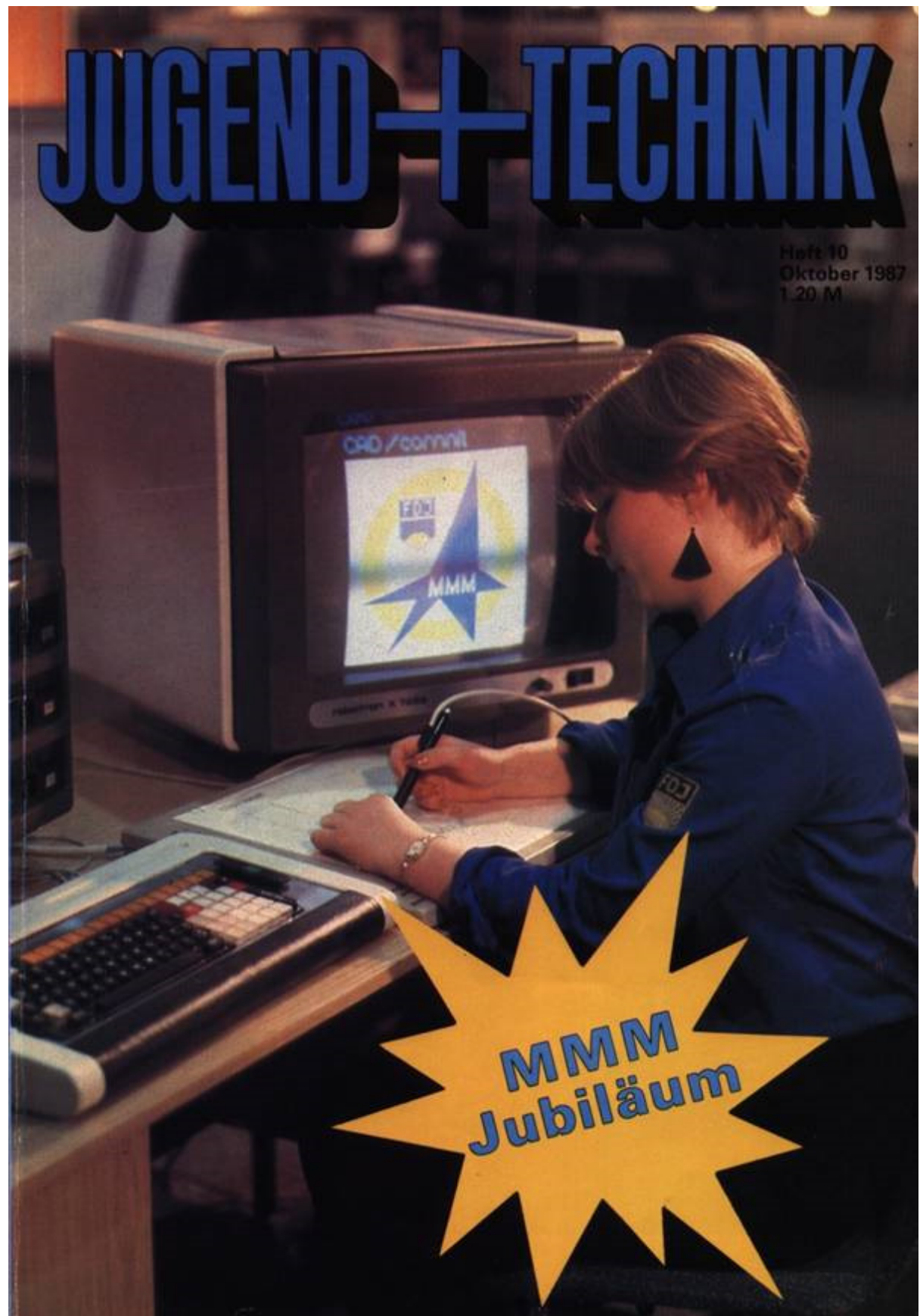

Note the importance of graphics: Color images at a time when color was exceptional! The artistic representation is peace and working class (blue), hope (stars) and mind (yellow). Note also the importance of youth and technology – and finally the computer operator is a woman. This is a very political and I think a progressive photograph. She appears to be using an early graphics tablet. Thus, at least some of the East German computer art clearly contained political content.
(source: http://robotron.informatik.hu-berlin.de/studienarbeit/files/literatur/pics/j_und_t.jpg)

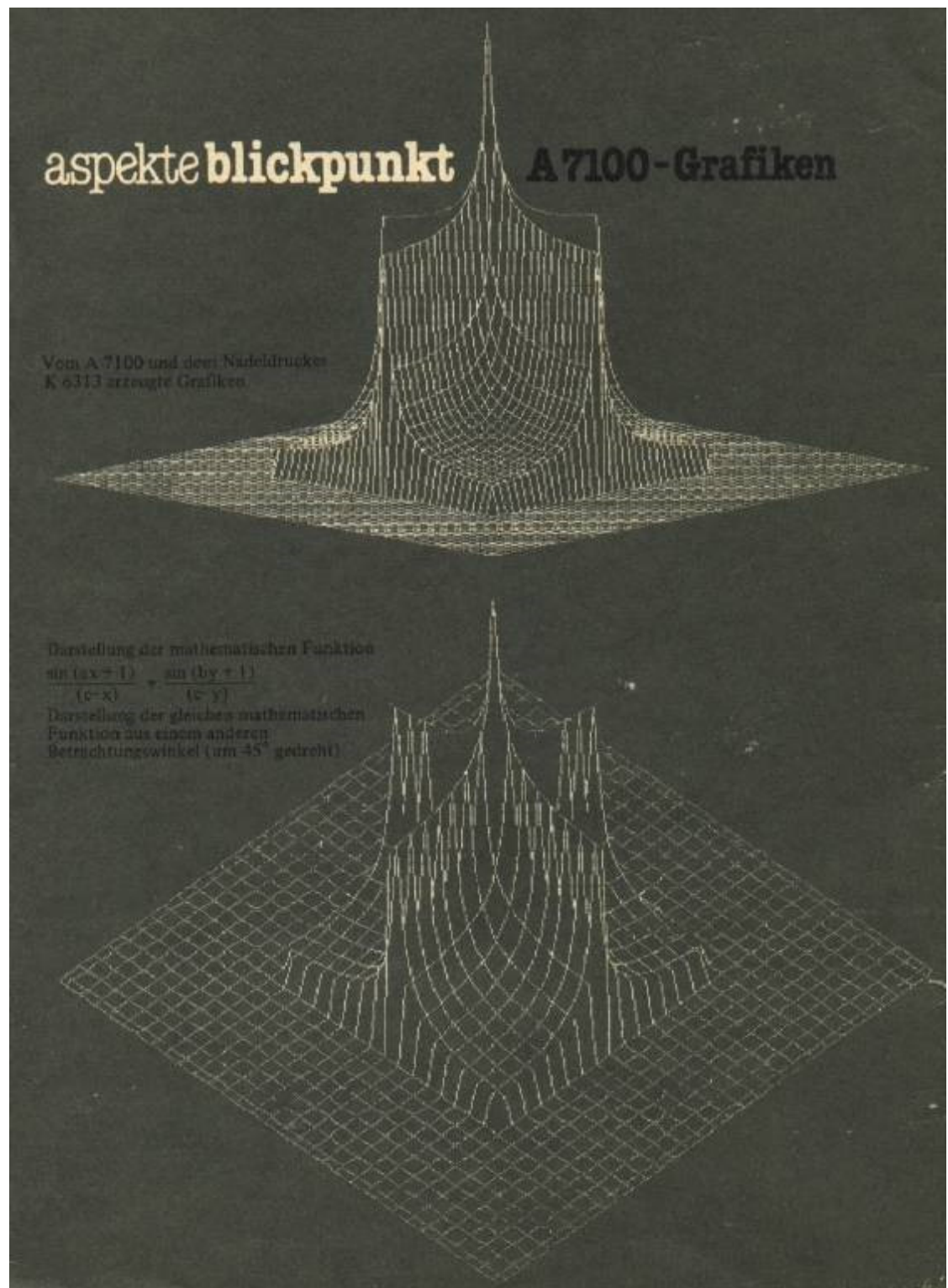

This graphic function reminds one of Mont St. Michel in France.

(Source: http://robotron.informatik.hu-berlin.de/studienarbeit/files/literatur/pics/edv-aspekte_02.jpg)

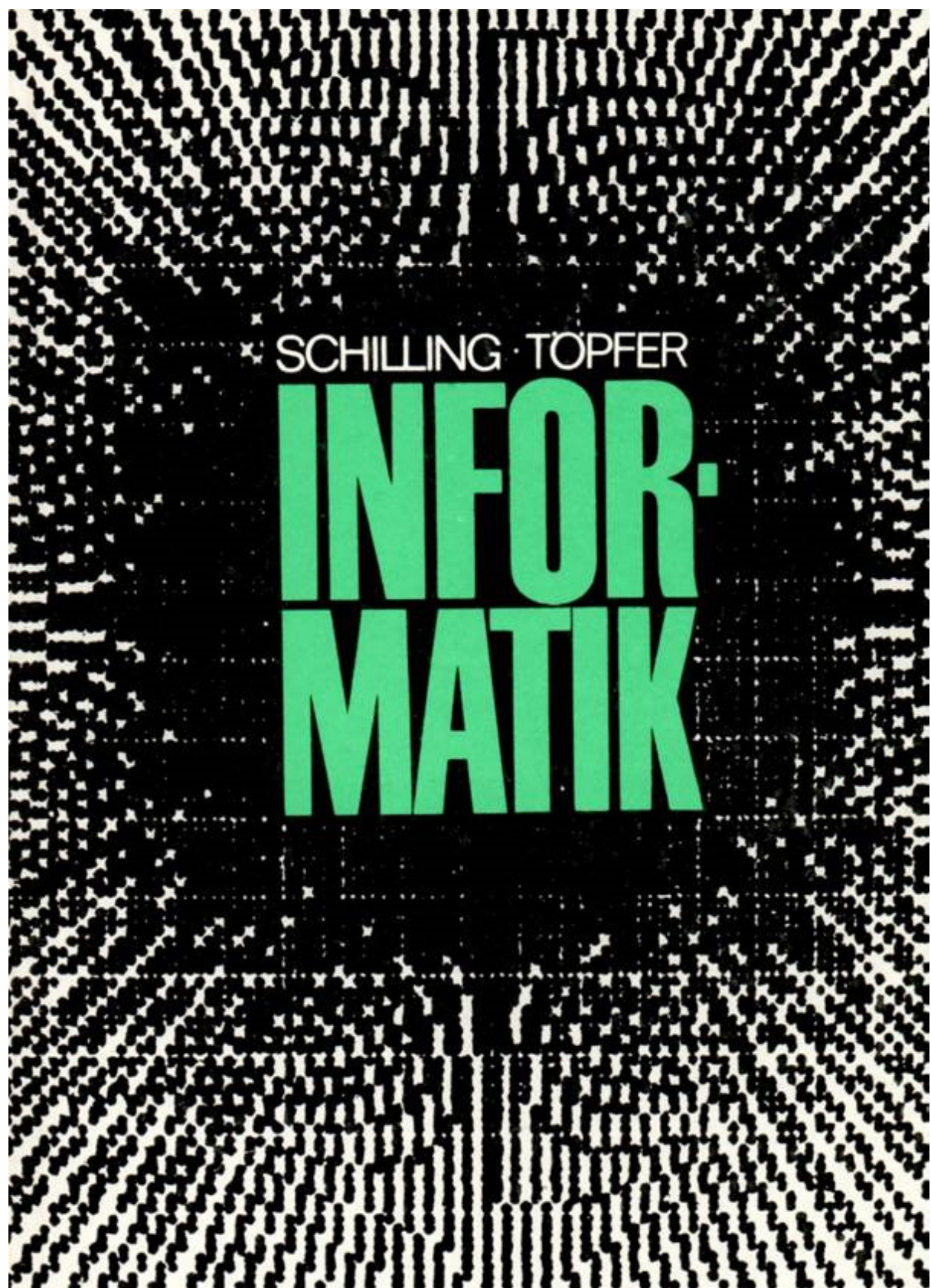

This is clearly a graphic function.

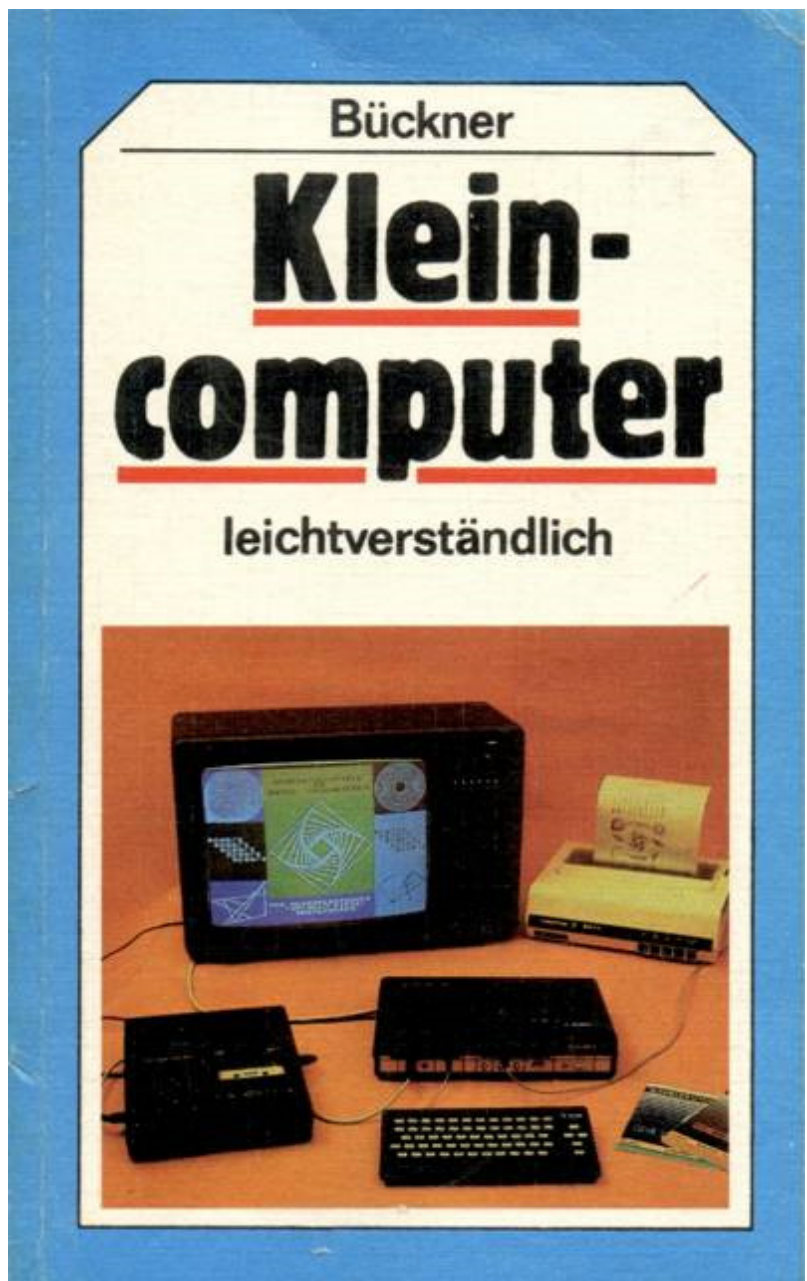

(source: http://robotron.informatik.hu-berlin.de/studienarbeit/files/literatur/pics/kleincomputer.jpg)

Note that this is obviously a book for home computer owners and not only for libraries and universities.
Again, notice the graphic representations (as well as the printer and cassette storage device!)

# Kleincomputer-Fibel

Jürgen Groh

```
299 REM   * GRAPH ZEICHNEN *
300 FOR X=-14 TO 14
310 Y=SGN(X):A=6
320 IF A$="INT" THEN Y=INT(X/5):A=3
330 PRINT AT(13-A*Y,19+X);"*"
340 NEXT X
398 :
399 REM  * PROGRAMM-ENDE *
400 PAUSE 50
410 CLS:PRINT
420 PRINT " NEUER PROGRAMMLAUF ? <J/N>"
430 Y$=INKEY$:IF Y$="" THEN 430
440 IF Y$="J" OR Y$="j" THEN 110
450 END
```

Akademie-Verlag Berlin

Note that the above BASIC program is the listing of a graphic function!

(source: http://robotron.informatik.hu-berlin.de/studienarbeit/files/literatur/pics/kleincomputer_fibel.jpg)

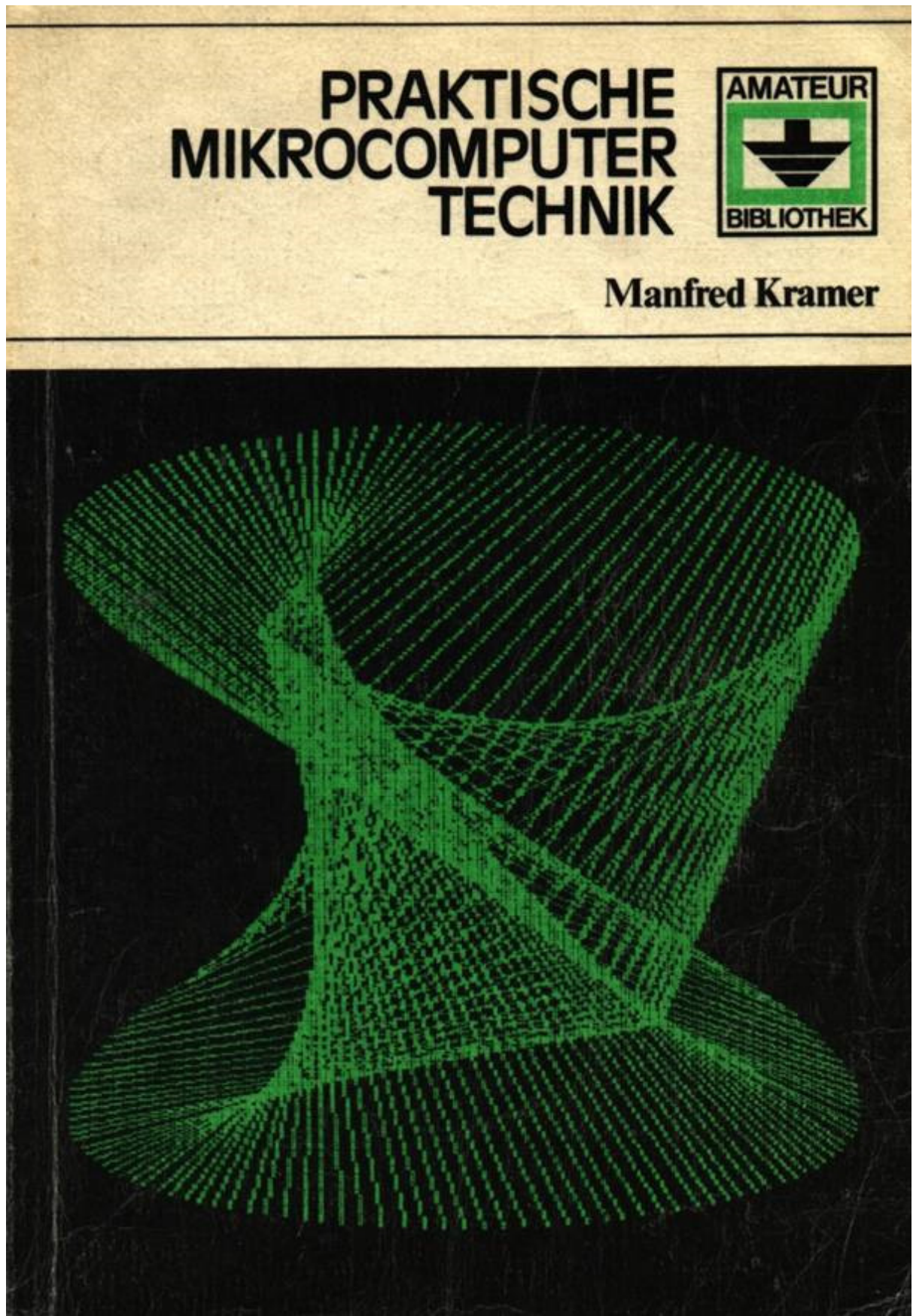

(source: http://robotron.informatik.hu-berlin.de/studienarbeit/files/literatur/pics/praktische.jpg)

Again, like the Yugoslav art this could be criticized as empty of class content. However we have seen that at least the Jugend and Technic cover illustrations computer generated art which clearly does portray class and gender struggle. These illustrations are also useful to demonstrate the level of technical achivement and scientific competence in East Germany.

## C. Conclusions

The technological capacities and achievements of East German computer science cannot be ignored. That these were also applied to artistic projects, at least in everyday life is also clear. Unfortunately, we see no movement like nove tendencije. So, while East Germany is unlike Yugoslavia and has not completely descended into neo-fascism the East German computer movement seems as yet to have made no permanent mark on history.

## D. Post-Wall Computer Art in (former) East Germany

As a divided country, Germany has perhaps the most interesting perspective on the East-West divide. Rudolf Kaehr draws lessons on computers and the arts. He argues that, nearly 10 years after the fall of the Berlin Wall, that lessons still need to be learned: According to Kaehr, the artist must become computer programmer and not merely be content with using tools created by others.[32]

Another artist who attempts to address East Germany using computer art is Eva Grubinger (born, 1970). An Austrian by birth she nonetheless works with East German political issues. In her "Operation R.O.S.A."[33] she presents a sculpture installation of a large child's toy as a foil for exposition of a complex, and perhaps fictional, account of how parents of a child use operant conditioning to "program" their child (the parents, incidentally, supposedly worked for the Robotron computer company in East Germany).
(example: http://www.nordschwarzwald.ihk24.de/PFIHK24/PFIHK24/servicemarken/regional/Anlagen/bilder-regional2002/rosa200.jpg)

Grubinger also works in "Anamorphosis" which is a computer technique for transforming a picture so that it can only be properly observed from one point. (example: http://www.flexibel.org/02_caught_in_flux/e_caught_in_flux_set.html ) She gives the audience directions for making this type of art. (See, http://www.flexibel.org/04_anamorphose/e_anamorphose_set.html)
Interestingly, she also develops "do it yourself art" – including a see-through bikini. (See, http://www.thing.at/netzbikini/)

## III. Software Art in the former Soviet Union

Though the history of Yugoslavian and East German computer art reveals that computer art was in fact a hot topic in East as in West the fact is that finding examples of computer art from the former U.S.S.R. seems difficult. However more recent anti-war examples of computer art are available from the Russian Federation. The defeat of the Soviet Union and its failed imperialist venture in Afghanistan – the Soviet Vietnam – led to great disillusionment. Artistically, this sadness expresses itself in the anti-war Net-Art of Olilia Lialina. Her artwork is innovative and essentially speaks for itself with a haunting message of love, fear, separation, and opposition to war. Her most famous work "My Boyfriend Came Back From the War" was made in 1996 using the then new technology of frames with black and white images. That media was appropriate because the topic was somber and the images would download quickly. Lialina and others, have since done reprises of the work in Flash, quicktime, (both in black and white and in color) and even as a downloadable "shooter" using Castle Wolfenstein – (also in Black and White).

The artist:
http://www.heise.de/tp/english/inhalt/on/5819/5819_1.jpg
http://sunsite.cs.msu.su/wwwart/aliona/full.htm
"My boyfriend came back from the war"
http://myboyfriendcamebackfromth.ewar.ru/

**Conclusions**

In conclusion, the attempt to achieve a classless society – a society where people would not be discriminated on the basis of race, sex, gender or caste – failed. It failed due to capitalist restoration resulting from revisionism. Any scientific consideration of Soviet economic performance prior to the revisionist Khruschev or in Maoist China reaches the result that both Stalin and Mao significantly extended the life expectency of their fellow citizens. The scientific achievements of the U.S.S.R. in Space (Sputnik), computer science and avionics were the result of careful planning and hard work. Because of the political failure of the the successors of the revolutionaries a bourgeoisie rose within the CPSU and the CPP. Consequently, capitalist restoration occurred throughout virtually all nominally communist countries. This material condition alone meant that any effort by artists, no matter how heroic, sincere, or hardworking would be doomed to a certain obscurity.

But social progress is not a matter of overnight change. It is rather a matter of generational changes. Today's reactionary may be tomorrows revolutionary. Progress means taking more steps forward than backward. Revisionism in Yugoslavia most acutely but also in the USSR and the PRC has resulted in a giant step backward toward the barbarism of nationalism, attendant genocide, and the risk of world war. Human rights since 1980 globally have in fact deteriorated. Economic inequality has grown. Racial discrimination and neo-fascism are also rising stars of the world we live in. But progress means to scientifically study one's actions. To repeat that which is succesful and correct one's errors. It was unthinkable in 1980 that the Soviet empire would implode. It is unthinkable today that capitalism will overextend itself in protracted wars for the resources on which it is dependant. But that is exactly what is happening. Revisionism is one step backward and given imperilaism another generation to kill the poor. It may provoke reaction and combine with capitalist resource dependancy to allow progressive forces to send society two steps forward at some point in the future. One thing is clear: a world divided by income, where torture and assassination are commonplace as instruments of foreign policy and constant wars for resources the norm is as undesirable as the pollution and racist poison which is now tolerated by too many. Eith the predictions of Marxism, that capitalism inevitably results in monopoly, war, and worsening cyclical crises are true or false. The world has had two great economic crises and two world wars to see this fact. If it requires a third world war to be convinced it may not survive its own scepticism.  The promise of Marxism is to make the transition from a market economy to a command economy and ultimately to communism as smooth as possible. But artists, when working to reduce human violence, must beware of revisionism's sugar-coated bullets. Whether the work of New Tendencies will one day redeem itself despite the political errors that allowed it to occur in the first place remains to be seen. To grow one hundred flowers, one must also remove revisionist weeds through political education, careful study, and self-criticism.

---

[1] "The history of avant-garde media art using moving pictures as the means of expression in Central and Eastern European countries is over 70 years old. Obviously, such experiment in the media artistic creation were dominated by film people, who often referred in their work to their experiences in the field of photography. In the first half of the 1970's video art began to develop parallel to film experiments… In recent years another transformation has been observed, resulting in the increasing interest of both artists and their audiences in interactive media art, placing its subjects in virtual reality and employing the communication potential of Internet." Ryszard W. Kluszczynski, *The Past and Present of (Multi)Media Art in Central and Eastern European Countries - An Outline* International Contemporary Art Network
 *(2003)* http://ican.artnet.org/ican/text?id_text=2

[2] Id.

[3] MIM, Revolutionary Definitions, (2003) http://www.etext.org/Politics/MIM/wim/revdefs.html

[4] MIM, Revolutionary Definitions (2003) http://www.etext.org/Politics/MIM/wim/revdefs.html

[5] Mao, U.S. Imperialism Is the Most  Ferocious Enemy of the World's People in  (1964)

[6] Mao,  "On Krushchov's Phoney Communism and Its Historical Lessons for the World: Comment on the  Open Letter of the Central Committee of the CPSU (IX)," Foreign Languages Press, Peking, July 14, 1964.

[7] "In 1953-1954 I spoke out [against reconciliation with Tito's] Yugoslavia at the Politburo. No one supported me, neither Malenkov nor even Kaganovich, though he was a Stalinist! Khruschev was not alone. There were hundreds and thousands like him, otherwise on his own he would not have gotten very far. He simply pandered to the state of mind of the people. But where did that lead? Even now there are lots of Khruschevs. . .

"Tito is now [in the 1970s] in a difficult situation. His republic is going under, and he will have to grab onto the USSR for dear life. Then we shall be able to deal with him more firmly.

"Nationalism is causing him to howl in pain, yet he himself is a nationalist, and that is his main defect as a communist. He is a nationalist, that is, he is infected with the bourgeois spirit. He is now cursing and criticizing his own people for nationalism. This means that the Yugoslav multinational state is breaking up along national lines. It is composed of Serbs, Croatians, Slovenes, and so forth.

"When Tito visited us for the first time, I liked his appearance. We didn't know everything about him at the time. . . .

Molotov's criticism was refined: He noted "There are many people worse than Tito and that "Tito is not an imperialist, he is a petty-bourgeois, an opponent of socialism. Imperialism is something else again."

Note: Albert Resis intro. & ed., Molotov Remembers: Inside Kremlin
Politics, Conversations with Felix Chuev (Chicago: Ivan R. Dee, 1993), pp.
83-4. Quoted from,
http://www.etext.org/Politics/MIM/classics/wetoldyouso/text.php?mimfile=molotovontito.txt
[8]
    "Later Mao was to write a succinct summary of the material roots of restoration of capitalism that pretty much sums up MIM's cardinal questions: In *Is Yugoslavia a Socialist Country*, Mao said, 'It shows us that not only is it possible for a working-class party to fall under the control of a labour aristocracy, degenerate into a bourgeois party and become a flunkey of imperialism before it seizes power.' Furthermore, Mao said, 'Old-line revisionism arose as a result of the imperialist policy of buying over and fostering a labour aristocracy. Modern revisionism has arisen in the same way. Sparing no cost, imperialism has now extended the scope of its operations and is buying over leading groups in socialist countries and pursues through them its desired policy of 'peaceful evolution.'' Hence, Mao always said the question of labor aristocracy is linked to the question of the restoration of capitalism. For a supposed Maoist to ignore the 'labor aristocracy' of the imperialist countries is revisionism. For people to talk about upholding the Cultural Revolution and opposing Soviet revisionism without opposing the labor aristocracy as enemy is just pure hogwash."
MIM, MC5 Book Review, From Trotsky to Tito, by James Klugmann
London: Lawrence & Wishart Ltd., 1951, 204pp. (June 28, 2002 ).
http://www.etext.org/Politics/MIM/bookstore/books/europe/klugmannyugo.html
[9]
    "From the mid-fifties studios in the various republics developed their own ideas and the directors" Heiko Daxl, *Film And Video-Art in Croatia Fragmentary Sketches of a History and a Description of the Status Quo*, August 1993. Id.

[10]
    "From the 1960s through the 1980s, counter-revolutionaries held up Yugoslavia as an example to sap the will of followers of Marx and Lenin and to serve as a half-way house for people who might otherwise make the leap to Marxism-Leninism-Maoism from capitalism. Yugoslavia was 'market socialism' with 'autonomous' enterprises. Noticeably missing was a Mao or Stalin-style central economic power. Yet, when all semblance of Stalin's influence was dead in the Soviet Union and the Gorbachev/Yeltsin bourgeoisie came out in the open, Yugoslavia fell apart most violently of all. That genocidal violence came as no surprise to us followers of Marx, Engels, Lenin, Stalin and Mao, because the material basis and guiding star of Yugoslavia's people had been set at the local level. The slogan, 'Think globally, act locally' can be an example of a profound mistake in this direction. What economic 'autonomy' did was to encourage the provinces to think at the expense of each other. That created the material basis for a narrowly provincial war." MIM, Godard's Maoist phase, 2003
http://www.etext.org/Politics/MIM/movies/review.php?f=long/godardmaoist.txt

[11]
    "Yugoslavia's break with the political doctrine of the Soviet Union under Stalin in 1948 lead to a policy of moderate Communism, an opening to West and to block neutralism. In the same way, areas of freedom were created permitting the following generations to gather experience in the countries of Western Europe and North America. Film studies, which officially didn't exist, consisted of endlessly watching films. The first important films of this time are documents of an uspurge from which many new things were to come."
Heiko Daxl, *Film And Video-Art in Croatia Fragmentary Sketches of a History and a Description of the Status Quo*, August 1993
http://home.snafu.de/mediainmotion/1994/crovideo.html
[12]
    Id.
[13]
    Id.
[14]
    Id.
[15]
    Museum of Contemporary Art, Zagreb, *New Tendencies*,
http://mama.mi2.hr/alive/eng/tendencije.htm.
[16]
    Id.
[17]
    Id.
[18]
    Darko Fritz, from the correspondence with Vuk Ćosić, February 2000.

[19] Statement by the collective Anonima in May 1968, catalogue 'tendencije 4' (1968 - 69), Zagreb, 1970.

[20] "… But the machines already approached the man, faster than the man approached the machines. …" Abraham A. Moles, introductory speech at the conference Computers and Visual Research, Zagreb, 1968., Bit International no. 2, 1968.

[21] "… this exhibition should not be understood as a domination of technology, but rather as an effort to overcome the new technology and use it to achieve new results in the field of visual." Boris Kelemen: Computer and visual research, catalogue 'tendencije 4' (1968 - 69), Zagreb, 1970.

[22] "A great deal of computer art embodies the limitation of existing techniques. The aesthetic demands of artists necessarily lead them to seek an alliance with the most advanced research in natural and artificial intelligence." Gordon Hyde, Jonathan Benthall, Gustav Metzger: Zagreb Manifesto, 1969, Bit International: Dijalog sa strojem, 1971.

[23] "if the households are going to be connected through the television screens with the central computer units, as it is nowadays the case with the phones, then nothing will stand in the way of the possibility to present computer graphics by means of a screen. That possibility seems today utopian …" Herbert W. Franke: Društveni aspekti kompjutorske umjetnosti [Social aspects of computer art], 1969, Bit International: Dalogue with the machine, 1971; "T.V. will be overshadowed by a C.V. (Computer vision) system combining and extending the present features of both computer and television systems removing the barrier of non-participation by the public. With increased free time, greater interest and activity we will be able to enjoy, and development of the arts and new tendencies should be in that direction." Petar Milojević: xxx, Bit International: : Dalogue with the machine, 1971.

[24] Gustav Metzger, exposition at the conference Computers and Visual Research, Zagreb, 1969., Bit International: Dijalog sa strojem, 1971.

[25] See, http://mama.mi2.hr/alive/eng/popis.htm

[26] Boris Kelemen: Computer and visual research, catalogue 'tendencije 4' (1968 - 69), Zagreb, 1970."

[27] See, http://www.ljudmila.org/~vuk/ascii/film/

[28] „Ein Problem war die Einlesbarkeit der ausgestrahlten Programme. Mit Hilfe der Studiotechnik des Rundfunks wurde die Erprobung durchgeführt. Die Programme wurden vom KC 85/3 auf Studioband (38 cm/s, mono, zweikanalig) aufgezeichnet, ausgestrahlt und dabei auf Kassette mitgeschnitten. Die Einlesbarkeit der Programme von Band in den Rechner war zufriedenstellend."
Tom Schnabel, *Computersendungen im Rundfunk der DDR,* http://robotron.informatik.hu-berlin.de/studienarbeit/files/hilfreiche_geister/voelz/rundfunk.html

[29] Id, http://robotron.informatik.hu-berlin.de/studienarbeit/files/literatur/literatur.html

[30] Id., http://robotron.informatik.hu-berlin.de/studienarbeit/files/rueckblick.html

[31] See, Tom Schnabel, http://robotron.informatik.hu-berlin.de/studienarbeit/files/software/software.html

[32] About the Art of Programming Art
Rudolf Kaehr, *Academy of Media Arts Cologne*, WEB: www.techno.net/pcl, Email: kaehr@khm.de "Self-modifying media lectures": http://smml.khm.de (1998)

[33] "Operation R.O.S.A.", (für abkürzungsversessene Zeitgenossen: "Rechnergestützte Orientierung der Selbstausbildung") besteht aus einer Skulptur und einem Video. Die Arbeit thematisiert, wie die führenden Technologien und Glaubenssysteme - Kapitalismus, Sozialismus, Technologie und Religion - das Individuum und (die Sprache) der Objekte beeinflussen können. Man sieht eine Skulptur - eine Riesen-Kinderrassel sowie das Video eines Kleinkindes, welches mit dieser Rassel spielt und hört aus dem "off" die Stimme der Figur einer jungen Deutschen namens Rosa, die in einem Monolog darlegt, warum sie eine existenzielle Krise durchlebt. Sie erzählt zunächst vom utopischen Traum ihrer Eltern - beides Wissenschaftler, angestellt bei der damaligen DDR-Computerfirma Robotron -, aus der Synthese von Kapitalismus, Individualismus und Sozialismus und durch das psychologische Verfahren der sogenannten "operativen Konditionierung" eine "bessere Welt" schaffen zu wollen.
http://www.nordschwarzwald.ihk24.de/PFIHK24/PFIHK24/produktmarken/index.jsp?url=http%3A //www.nordschwarzwald.ihk24.de/PFIHK24/PFIHK24/servicemarken/regional/kultur/projekt-rosa.jsp